\RequirePackage{fix-cm}
\documentclass{svjour3}                     

\smartqed  

\usepackage{graphicx}
\usepackage{natbib}
\usepackage{multirow}
\usepackage{amssymb,amsmath}

\begin{document}

\title{Chiral Polymerization in Open Systems From Chiral-Selective Reaction Rates
}

\author{Marcelo Gleiser       \and
        Bradley J. Nelson \and 
        Sara Imari Walker
}


\institute{M. Gleiser \at
              Department of Physics and Astronomy, Dartmouth College
Hanover, NH 03755, USA \\
              \email{mgleiser@dartmouth.edu}  
           \and
           B. Nelson \at
           Department of Physics and Astronomy, Dartmouth College
Hanover, NH 03755, USA
           \and
           S. Walker \at
              NASA Astrobiology Institute, USA\\
              BEYOND: Center for Fundamental Concepts in Science, Arizona State University, Tempe, AZ\\
}

\date{Received: date / Accepted: date}

\maketitle

\begin{abstract}
We investigate the possibility that prebiotic homochirality can be achieved exclusively through chiral-selective reaction rate parameters without any other explicit mechanism for chiral bias. Specifically, we examine an open network of polymerization reactions, where the reaction rates can have chiral-selective values. The reactions are neither autocatalytic nor do they contain explicit enantiomeric cross-inhibition terms. We are thus investigating how rare a set of chiral-selective reaction rates needs to be in order to generate a reasonable amount of chiral bias. We quantify our results adopting a statistical approach: varying both the mean value and the rms dispersion of the relevant reaction rates, we show that moderate to high levels of chiral excess can be achieved with fairly small chiral bias, below 10\%. Considering the various unknowns related to prebiotic chemical networks in early Earth and the dependence of reaction rates to environmental properties such as temperature and pressure variations, we argue that homochirality could have been achieved from moderate amounts of chiral selectivity in the reaction rates.
\end{abstract}

\section{Introduction}
\label{intro}
The origin of life on Earth is a central scientific question of our time \protect\citep{origin,orgel}. A possible way of organizing its many facets is to separate it into three basic aspects, all still unknown: when did it happen; where did it happen; and how did it happen. The ``when'' question is, to a certain extent, the best understood. We can put a lower limit on the origin of life on Earth at about 3.5 billion years ago (BYA) \citep{timing}. Possible evidence of earlier life, dated at 3.8 BYA, remains controversial \citep{Arrhenius}. Of course, it is also possible that life emerged more than once before this time, although if it did it was probably made extinct quickly due to the unstable conditions of a crustless primeval Earth. Another intriguing possibility is that of a shadow biosphere, which has hitherto remained undetected \citep{Benner,Davies1}. At about 3.9 BYA, the Late Heavy Bombardment caused by the displacement of outer planets, constitutes a serious obstacle against any long-term survivability of earlier life forms \citep{bombardment}. 

The ``where'' question has many possible answers, ranging from shallow water pools and hot water vents to clay and mineral surfaces \citep{where}. It is also speculated that life may have originated in outer space, possibly Mars \citep{Marslife}, and was transplanted to Earth via meteoritic impacts \citep{Davies1996,Panspermia1,origin}. If not life itself, at least some of its basic organic ingredients may indeed have come from outer space, possibly seeding early life here \citep{sagan,pizzarello,Panspermia2}. 

The ``how'' question remains quite open and it may never be answered precisely, as it depends on the details of a prebiotic biochemistry we may never completely know. However, the uniformity in the biochemistry of all living organisms strongly suggests that all life descended from a last universal common ancestor (LUCA), that lived 3.2-3.8 BYA and had a level of complexity similar to that of a simple modern bacterium \citep{orgel}. To ``explain'' the origin of life requires us to explain how abiotic organic compounds combined to generate the LUCA, at present a major challenge. 

A key aspect of all terrestrial life is its biomolecular homochirality \citep{chiralreview1,chiralreview2}: the vast majority of amino acids comprising living creatures are levorotatory ($L$), while the majority of sugars are dextrorotatory ($D$). Given the link of present and past life to the LUCA, we should thus expect that the LUCA displayed the same chiral signature as modern life. To explain the origin of life we must also therefore explain the origin of homochirality. Although it is possible that biomolecular homochirality evolved during, or even after, prebiotic chemistry reached a more complex stage as, say, when the first nucleic acids were being formed \citep{RNAworld}, it is more likely  that homochirality is a precondition to life. Accordingly, we will assume here that chiral selectivity played a key role as prebiotic biochemistry began to evolve, setting the conditions for the first living forms to emerge. Within this framework, to understand the ``how'' question we must first understand the homochiral question: essentially, we must understand how the first homochiral polymers formed under prebiotic conditions.

Over the past decades, several explanations have been offered to account for the homochirality of living organisms \citep{chiralreview1,chiralreview2}. All of them assume that the essential reactions leading to homochirality are highly nonlinear and unstable to growth, so that a small excess of one handedness can be efficiently amplified to full homochirality. There are thus two basic aspects of the homochiral question: first, the cause for the initial excess of one handedness over the other; and second, the amplification mechanism.

Causes for the initial chiral excess can be classified as endogenous or exogenous. By endogenous causes we mean those that are intrinsic to the physics and/or chemistry of the reactor pool. For example, parity symmetry breaking in the weak nuclear interactions has been proposed as a possible source of the initial homochiral excess \citep{yamagata, salam, kondepudi}, although the numbers obtained in quantum computations using perturbed Hamiltonians are exceedingly small \citep{quack}. 
In contrast, exogenous sources are caused by outside influences. Examples of exogenous sources of initial chiral bias include circularly-polarized UV light from active star formation regions \citep{bailey}, and deposits of chiral compounds by meteoritic bombardment \citep{pizzarello, glavin}. Another possibility of an exogenous influence is that chirality is induced by a clay template, such as in recent results for n-Propyl NH$_3$Cl vermiculite clay gels \citep{Fraser}, which exhibit chiroselective absorption of amino acids.

At this point, the debate is still open as to whether any of these possible sources of chiral bias was present at prebiotic times and, if it was, whether it was efficient or necessary. Gleiser, Walker, and Thorarinson have recently argued that even if an initial chiral excess was produced by any of these (or other) mechanisms, environmental disturbances could have erased it, restoring either achiral conditions or switching chirality in the opposite direction: chiral selection might be the result of a series of stochastic reversals prompted by large temperature and/or density fluctuations, or other possible environmental effects \citep{punctuated}. Similar conclusions were reached recently by \citet{Hochberg}. Accordingly, the question of biological homochirality cannot be separated from the details of planetary history. The same holds for any other planetary platform displaying stereoselective chemistry.

The second basic aspect of the homochiral question focuses on the details of the amplification mechanism. Ever since the seminal work by Frank \citep{Frank}, most models attempt to describe how a small chiral bias can be efficiently amplified toward homochirality. Such models rely on a combination of two properties: the reactions must be autocatalytic, thus guaranteeing the nonlinear behavior needed for amplification, and they must display enantiomeric cross-inhibition, that is, enantiomers of opposite chirality can combine, suppressing the growth of homochiral chains \citep{Sandars, wattis, saito}. As can be seen in the Sandars model \citep{Sandars}, this inhibition may occur by suppressing the enzymatic activity of homochiral chains in the reactor pool. Within this framework, one must still appeal to a mechanism for generating an initial excess. Traditional approaches along these lines have held that the initial symmetry breaking is endogenous, occurring at the level of small stochastic fluctuations in the concentrations of left-- and right--handed monomers, which must then be amplified by the combined effects of autocatalysis and cross-inhibition to yield an appreciable enantiomeric excess \citep{Sandars, gleiser_walker}. 
However, fluctuations between the concentrations of left-- and right-- handed enantiomers may be exceedingly small in a large reactor pool, thereby requiring strong autocatalytic feedback for amplification to occur, if at all. Given that autocatalytic feedback leading to significant chiral selection has only been observed in a few abiotic systems \citep{saito}, and has not yet been observed in a prebiotically relevant scenario, it is prudent to consider other possibilities for the onset of chiral selection in a prebiotic environment.

In this work we therefore explore an alternative scenario for the generation of a chiral excess in open polymerization systems without invoking autocatalysis or enantiomeric cross-inhibition. Instead, we propose that reasonable amounts of chiral excess can be achieved solely through chiral-selective reaction rates. In order to sample a large number of possible values for the reaction rates, we present a statistical analysis whereby each run represents a potential prebiotic chemistry with fixed values of the reaction rates.  For each prebiotic scenario the rates are picked from a sample of Gaussian-distributed values. Although a large statistical sampling of the systems under study would on average be racemic,  we show that moderately small chiral-selective variations in reaction rates may lead to significant chiral excess in a steady-state. Within a prebiotic scenario on early Earth, we might imagine that such chiral-selective fluctuations in the reaction rates were prompted by environmental disturbances or by  still unknown details of the prebiotic chemistry. We propose a few possibilities, although the strength of our statistical analysis is its independence from the specific mechanism generating the chiral excess in the reaction rates.Thus, we will explore the possibility that terrestrial biomolecular homochirality might have arisen without the need for any of the conventional mechanisms for chiral biasing or amplification. For illustrative values of the relevant reaction rates, we will place lower bounds on the amplitude of fluctuating reaction rates needed for substantial chiral bias.

\section{Chiral Polymerization Model}
\label{cpm}

Consider a polymerization reaction network where activated levorotatory and dextrorotatory monomers ($L_1^*$ and $D_1^*$, respectively), can chain up to generate longer homochiral or heterochiral molecules. (Here, ``activated'' refers to these monomers being highly reactive due to energy input into the reactor pool, as in the APED model of Plasson {\it et al.}, proposed as an alternative to Frank-like models \citep{Plasson}). The model reactions include deactivation of activated monomers,
\begin{align}
L_1^* \stackrel{h_L}\rightarrow L_1  && D_1^* \stackrel{h_D}\rightarrow D_1 
\end{align}
and the polymerization reactions,
\begin{align}
L_1^* + L_i \stackrel{a_L}\rightarrow L_{i+1}  && D_1^* + D_j \stackrel{a_D}\rightarrow D_{j+1} \\
L_1^* + D_j \stackrel{a_L}\rightarrow M_{1j}  && D_1^* + L_i \stackrel{a_D}\rightarrow M_{i1} \\
L_1^* + M_{ij} \stackrel{a_L}\rightarrow M_{i+1j}  && D_1^* + M_{ij} \stackrel{a_D}\rightarrow M_{i j+1}
\end{align}
where $h_{L(D)}$ is the deactivation rate for $L(D)$-monomers and $a_{L(D)}$ is the polymerization rate for adding activated $L(D)$-monomers to a growing chain. Here $L_i$ and $D_j$ denote homochiral polymers of length $i$ and $j$ respectively, and $M_{ij}$ denotes polymers of mixed chirality consisting of $i$ $L$-monomers and $j$ $D$-monomers. Note that this reaction network is substantially different from the usual Frank-like model for chiral symmetry breaking \citep{Frank,Sandars}: there is no autocatalysis (either explicit or through enzymatic activity of homochiral chains) or enantiomeric cross-inhibition present. We note that an alternative copolymerization model with two chemically distinct monomers has been recently proposed by Wattis and Coveney \citep{wattis07}.

What kinds of situations could lead to chiral-selective values for the reaction rates? It is well-known that the temperature dependence of reaction rates is determined by the Arrhenius factor, $k \sim \exp[-E_a/RT]$, where $E_a$ is the activation energy related to departures from equilibrium populations of molecules. Consider, for instance, a situation where levorotatory and dextrorotatory rates have {\it different} temperature dependencies, parameterized by chiral-specific activation barriers as $k_{L(D)} = \exp[-E_{L(D)}(T)/RT]$. For example, we could model this dependence as $E_{L(D)} = -c_{L(D)}T^2 + d_{L(D)}$, where the four constants $c_{L(D)}$ and $d_{L(D)}$ are positive. If, say, $c_L > c_D$ and $d_L = d_D$, we see that for all temperatures (for $T<T_c^2 = d_D/c_D$) dextrorotatory reactivity is enhanced. A different realization would have correspondingly different constants or a different dependence of the activation energy with temperature. Each choice would correspond to an independent prebiotic scenario. (We should consider this model only as a suggestive mechanism whereby chiral selectivity could occur through the reaction rates.)

Given that the details of prebiotic chemistry remain elusive, here we are considering that, in principle, reaction rates may be chiral-selective. However, instead of assigning {\it ad hoc} values for the rates of opposite chirality, we will conduct a statistical analysis whereby, for different numerical experiments--that is, different realizations of prebiotic scenarios--the values for $L$ and $D$ reactions rates will be allowed to randomly fluctuate about a given mean. In other words, each time we solve the coupled nonlinear ordinary differential equations describing the reactions, we will pick a set of random, Gaussian-distributed values for the four reactions rates $h_{L(D)}$ and $a_{L(D)}$. This heuristic approach will allow us to investigate the range in the values of the reaction rates that generates a substantial departure from racemic results. As we will see, modest values for the fluctuation range for $L$ and $D$ reaction rates will suffice to generate substantial chiral bias. Although we do not model any specific chemical system, we expect that our general results will be applicable to a range of experiments probing the rate-dependence of polymerization models and its impact on chiral symmetry breaking.

In practice, we will consider pairs of randomly-distributed reaction rates $(h_L,h_D)$ and $(a_L,a_D)$ taken from a Gaussian distribution with a racemic mean
\begin{eqnarray}
\label{mean}
\langle h_L \rangle  &=& \langle h_D \rangle = \bar h\\ \nonumber
\langle a_L \rangle  &=& \langle a_D \rangle = \bar a,
\end{eqnarray}
and a two-point correlation
\begin{eqnarray}
\label{rms}
\langle h_{L(D)} h_{L'(D')}\rangle  &=& h_0^2 \delta_{LL'(DD')}\\ \nonumber
\langle a_{L(D)} a_{L'(D')}\rangle  &=& a_0^2 \delta_{LL'(DD')},
\end{eqnarray}
where $h_0$ and $a_0$ are the rms values, and $\delta_{xx'}$ is the Kronecker delta. The angled brackets mean an average over $N$ experiments (an ensemble average), with $N$ large. For example, the normalized probability $P(h_L)$ that the $L$-hydrolysis rate will obtain a value $h_L$ is 
\begin{equation}
\label{gaussian}
P(h_L) = \frac{1}{\sqrt{2\pi h_0^2}}\exp[-(h_L - \bar h)^2/2h_0^2].
\end{equation}

For the above system, the rate equations for the various concentrations are written as (from here on, quantities in capitals denote the concentrations of the different reactants):
\begin{eqnarray}
\frac{d L_1^*}{dt} &=& S- h_L L_1^* - a_L L_1^* \left( \sum_{i=1}^\infty L_i + \sum_{j=1}^\infty D_j + \sum_{i=1}^\infty \sum_{j=1}^\infty M_{ij}  \right) - d L_1^*\\
\frac{d D_1^*}{dt} &=& S- h_D D_1^* - a_D D_1^* \left( \sum_{i=1}^\infty L_i + \sum_{j=1}^\infty D_j + \sum_{i=1}^\infty \sum_{j=1}^\infty M_{ij}  \right) - d D_1^*\\
\frac{d L_1}{dt} &=& h_L L_1^* - a_L  L_1 L_1^* -a_DL_1 D_1^* - d L_1\\
\frac{d D_1}{dt} &=& h_D D_1^* - a_LD_1 L_1^*-a_D D_1 D_1^*  - d D_1\\
\frac{d L_i}{dt} &=& a_L L_1^* L_{i-1}  - a_L L_i L_1^* -a_DL_i D_1^*  - d L_i~~~~ ( i  \geq 2)\\
\frac{d D_j}{dt} &=& a_D D_1^* D_{j-1}  - a_L D_j L_1^* -a_DD_jD_1^*  - d D_i~~~~ ( j \geq 2)\\
\frac{d M_{ij}}{dt} &=& a_L( L_1^* M_{i-1 j}-M_{ij} L_1^*) + a_D (D_1^* M_{i j-1} -M_{ij}D_1^*)  - d M_{ij}~( i,j \geq 1)
\end{eqnarray}

Note that for the open systems studied here we also include the additional rates $S$ describing the permeation of activated $L_1^*$ and $D_1^*$ monomers into the system, and $d$ describing the loss of all molecular species from the system. For closed systems $S = d = 0$. 

We can simplify the reactions above by defining
\begin{align}
{\cal L} = \sum_{i=2}^\infty L_i,~~~~~ {\cal D} = \sum_{j=2}^\infty D_j,~~~~~{\cal M} =  \sum_{i=1}^\infty \sum_{j=1}^\infty M_{ij}, 
\end{align}
and summing the reaction equations over $i$ and $j$.
Then the reaction--network, in the limit of infinite-length polymers, reduces to:
\begin{eqnarray} 
\frac{d L_1^*}{dt} &=& S- h_L L_1^* - a_L L_1^* \left(L_1+{\cal L}+ D_1+{\cal D} + {\cal M} \right) - d L_1^*;\\
\frac{d D_1^*}{dt} &=& S- h_D D_1^* - a_D D_1^*\left(L_1+{\cal L}+ D_1+{\cal D} + {\cal M} \right) - d D_1^*;\\
\frac{d L_1}{dt} &=& h_L L_1^* - a_L  L_1 L_1^* -a_DL_1 D_1^* - d L_1;\\
\frac{d D_1}{dt} &=& h_D D_1^*  -a_LD_1 L_1^* - a_D D_1 D_1^* - d D_1;\\
\frac{d {\cal L}}{dt} &=& a_L L_1^* L_1  - (a_D D_1^* + d) {\cal L};\\
\frac{d {\cal D}}{dt} &=& a_D D_1^* D_1  - (a_L L_1^* + d) {\cal D};\\
\frac{d {\cal M}}{dt} &=& a_L L_1^* (D_1 + {\cal D}) + a_D D_1^* (L_1 + {\cal L}) - d {\cal M}.
\end{eqnarray}

The quantity of interest is the enantiomeric excess [$ee(t)$], that is, the net chirality, which we define as

\begin{equation}
\label{ee_evol}
ee(t)=\frac{ [L_1^*(t)+L_1(t)+{\cal L}(t)]-[D_1^*(t)+D_1(t)+{\cal D}(t)]}{[L_1^*(t)+L_1(t)+{\cal L}(t)]+[D_1^*(t)+D_1(t)+{\cal D}(t)]}.
\end{equation}
Note that our definition of the enantiomeric excess implicitly assumes that we are only interested in pure homochiral polymers.
Given initial values for the concentrations, the equations describing their evolution can be solved and the net chirality computed as a function of time. We will choose the simplest possible initial conditions, setting all initial concentrations to zero, with the exception of $L_1^*(0)$ and $D_1^*(0)$, which we will set equal to each other so that the system starts with no chiral bias and no long polymer chains. We will also pick a constant value for the source for activated monomers, $S(t)=S.$ With this choice, we will be able to investigate under which conditions, that is, under which values for the reaction rates, the system engenders a net chirality as it evolves to a steady-state wherein all concentrations achieve a dynamic equilibrium so that their time derivatives vanish. We note that results were not very sensitive to the choice of $S$.

Before we show solutions to the system of equations, we eliminate the loss rate from the equations, dividing them on both sides by $d$. Since $d$ has dimensions of inverse time, this is equivalent to introducing the dimensionless time variable $\tau = td$ such that we can write the equations as,

\begin{eqnarray} 
\label{scaledreactioneqs1}
\frac{d L_1^*}{d\tau} &=& \sigma- \beta_L L_1^* - \alpha_L L_1^* \left(L_1+{\cal L}+ D_1+{\cal D} + {\cal M} \right) -  L_1^*;\\
\frac{d D_1^*}{d\tau} &=& \sigma- \beta_D D_1^* - \alpha_D D_1^*\left(L_1+{\cal L}+ D_1+{\cal D} + {\cal M} \right) -  D_1^*;\\
\frac{d L_1}{d\tau} &=& \beta_L L_1^* - \alpha_L  L_1 L_1^* -\alpha_DL_1 D_1^* -  L_1;\\
\frac{d D_1}{d\tau} &=& \beta_D D_1^*  -\alpha_LD_1 L_1^* - \alpha_D D_1 D_1^* -  D_1;\\
\frac{d {\cal L}}{d\tau} &=& \alpha_L L_1^* L_1  - (\alpha_D D_1^* + 1) {\cal L};\\
\frac{d {\cal D}}{d\tau} &=& \alpha_D D_1^* D_1  - (\alpha_L L_1^* + 1) {\cal D};\\
\label{scaledreactioneqs7}
\frac{d {\cal M}}{d\tau} &=& \alpha_L L_1^* (D_1 + {\cal D}) + \alpha_D D_1^* (L_1 + {\cal L}) -  {\cal M},
\end{eqnarray}
where we have introduced $\sigma\equiv S/d$, $\alpha_{L(D)}\equiv a_{L(D)}/d$, and $\beta_{L(D)}\equiv h_{L(D)}/d$. With this parametrization $\alpha_0 = a_0/d$, $\beta_0 = h_0/d$, and $\bar{\alpha} = \bar{a}/d$ and $\bar{\beta} = \bar{h}/d$ characterize the random distributions of reaction rates outlined in eqs. \ref{mean} and \ref{rms}. Thus, all reaction rates and dynamical time-scales are expressed in terms of the disappearance rate $d$. In particular, $\sigma$ gives the ratio per unit volume of entrance versus exiting of reacting materials to and from the reactor pool.

\section{Results}
\label{results}

In Figure \ref{ee_evol} we show the behavior of the time evolution of enantiomeric excess for several illustrative examples with $\sigma = 200$ and different choices for the reaction rates $\alpha_L$, $\alpha_D$, $\beta_L$ and $\beta_D$. On the left, we show results with $\beta_L = \beta_D=100$ fixed, and varying ratios of $\alpha_L/\alpha_D$. Shown are the results for $\alpha_L=5.0$, $\alpha_D=10.0$ ($\alpha_L/\alpha_D = \frac{1}{2}$) in blue, $\alpha_L=5.0$, $\alpha_D=15.0$ ($\alpha_L/\alpha_D = \frac{1}{3}$) in black, and $\alpha_L=5.0$, $\alpha_D=20.0$ ($\alpha_L/\alpha_D = \frac{1}{4}$) in red.  The system quickly reaches a steady state with net chiral excess $ee_{ss}\simeq 0.24$,  $ee_{ss}\simeq 0.38$, and  $ee_{ss}\simeq 0.48$, respectively (corresponding to $24\%$, $38\%$, and $48\%$ enantiomeric excess). On the right, we show results holding  $\alpha_L= \alpha_D=10.0$ fixed, and varying the ratio of $\beta_L/\beta_D$. Shown are the results for $\beta_L = 75$, $\beta_D = 25$ ($\beta_L/\beta_D = 3$) in blue,  $\beta_L = 100$, $\beta_D = 25$ ($\beta_L/\beta_D = 4$) in black, and $\beta_L = 125$, $\beta_D = 25$ ($\beta_L/\beta_D = 5$) in red. In this case the net chiral excess at steady-state is $ee_{ss}\simeq 0.38$,  $ee_{ss}\simeq 0.47$, and  $ee_{ss}\simeq 0.53$, respectively (corresponding to $38\%$, $47\%$, and $53\%$ enantiomeric excess). The choice of $\sigma$ is predicated by the value of $\beta_{L(D)}$, since for $\sigma \ll \beta_{L(D)}$ the suppression of activated monomers would be too effective. We investigated values for $\sigma=100$ and $\sigma=300$ but did not find any significant variation in the outcomes.

These test runs show that for differing left and right reaction rates a substantial amount of chiral excess may be reached at steady-state. Although for these examples the differences were fairly large, we note that if the chiral-selective changes in rates appear in exponential factors (as in the example above for temperature dependence), small changes in the parameters may generate fairly large changes in the resulting reaction rates. 

\begin{figure}[htbp]
\includegraphics[width=0.49\linewidth]{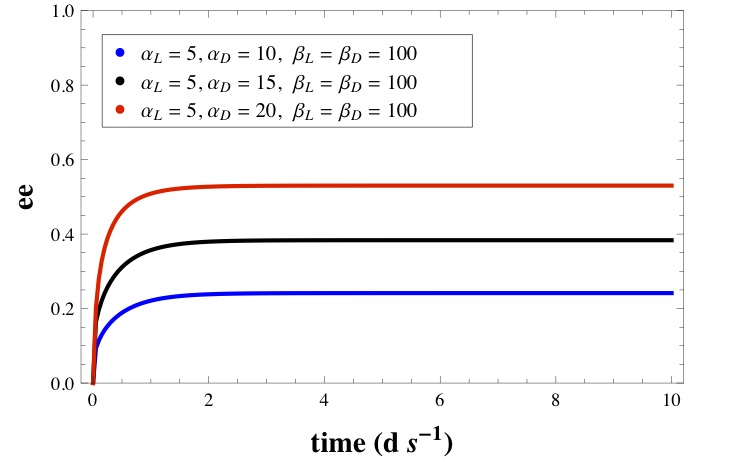}
\includegraphics[width=0.49\linewidth]{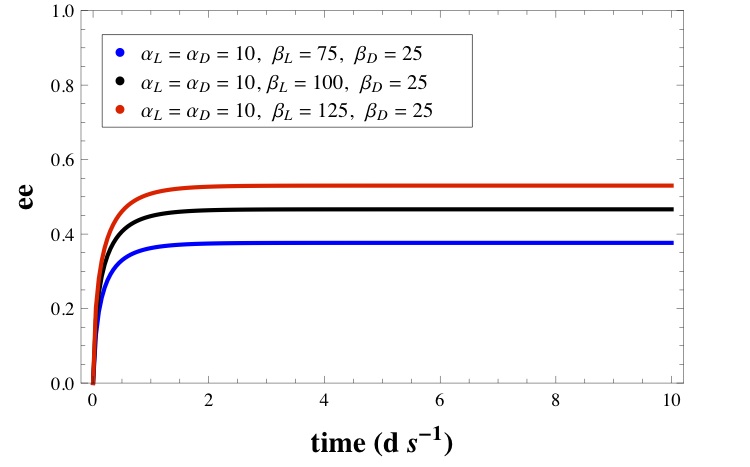} 
\caption{Time evolution of enantiomeric excess for illustrative examples of reaction rates with $\sigma = 200$. {\bf Left:} $\beta_L = \beta_D=100$, with varying ratios of $\alpha_L/\alpha_D$. Shown are $\alpha_L=5.0$, $\alpha_D=10.0$ ($\alpha_L/\alpha_D = \frac{1}{2}$) in blue, $\alpha_L=5.0$, $\alpha_D=15.0$ ($\alpha_L/\alpha_D = \frac{1}{3}$) in black, and $\alpha_L=5.0$, $\alpha_D=20.0$ ($\alpha_L/\alpha_D = \frac{1}{4}$) in red. {\bf Right:} $\alpha_L= \alpha_D=10.0$, with varying ratios of $\beta_D/\beta_L$. Shown are  $\beta_L = 75$, $\beta_D = 25$ ($\beta_L/\beta_D = 3$) in blue,  $\beta_L = 100$, $\beta_D = 25$ ($\beta_L/\beta_D = 4$) in black, and $\beta_L = 125$, $\beta_D = 25$ ($\beta_L/\beta_D = 5$) in red}
\label{ee_evol}
\end{figure}

In general, we note the reactions favor an $L$-excess for ratios $\alpha_L/\alpha_D < 1$ and/or $\beta_L/\beta_D > 1$ as demonstrated by the results shown in Figure 1. At first glance this may seem counterintuitive, particularly the relationship between $\alpha_L/\alpha_D$ and the net enantiomeric excess of homochiral polymers. Intuitively one would expect that increasing $\alpha_D$ (the polymerization rate of activated $D$-monomers) while holding $\alpha_L$ (the polymerization rate of activated $L$-monomers) fixed would result in a net excess favoring $D$, and not $L$ as observed. However, on closer inspection one must note that $\alpha_D$ describes the rate of attachment of activated $D$-monomers to growing $D$-- {\it or} $L$-- chains, leading to higher yields of $D$-enriched mixed polymers and a correspondingly lower yield of pure $D$-polymers. Likewise, increasing the deactivation rate of $L$-monomers  ({\it i.e.} $\beta_L$) and thereby increasing the ratio $\beta_L/\beta_D$ leads to fewer $L$-activated monomers for polymerization which decreases the yield of pure homochiral $L$- polymers, but also decreases the loss of $L$-monomers to polymers of mixed chirality, resulting in a net increase in the enantiomeric excess favoring $L$. Clearly, with the same values of $L$ and $D$ reaction rates, $ee=0$ for all $t$. We must thus examine how the results behave for statistical samplings of
$\alpha_L$, $\alpha_D$, $\beta_L $, and $\beta_D.$

It is also important to compare the production of homochiral and heterochiral polymers, since the na\"ive expectation is that the system will be swamped by heterochiral polymers. In Table \ref{homohetero_results}, we used the same reaction rates as in Figure \ref{ee_evol} to show that this is not the case. In fact, the ratio of homochiral to heterochiral polymers [defined as ${\cal R}=({\cal L} + {\cal D})/{\cal M}$] is fairly invariant with respect to changes in the reaction rate parameters (grouped in the Table as blocks of three rows each), suggesting the possible existence of attractors in the phase space of the reaction kinetics determining the equilibrium concentration of polymers.

\begin{table}[htdp]
\begin{center}
\begin{tabular*}{0.665\textwidth}{|c|c|c|c|c|c|c|c|} \hline
$\alpha_L$   &$\alpha_D$ &  $\beta_L$  & $\beta_D$ & $ee$ & ${\cal M}$ & ${\cal L} + {\cal D}$ & ${\cal R}$\\ \hline \hline
 5	&10	&100 &100 &0.242  & 49.154 & 8.328 & 0.170   \\ \hline
 5	&15	&100 &100 & 0.384 & 47.365 & 7.255 & 0.153 \\ \hline
 5	&20	&100 &100 & 0.479 & 46.345 & 6.630 & 0.143    \\ \hline \hline
 10	&10	&75	 &25   & 0.376 & 33.958 & 3.391   & 0.100 \\ \hline
 10	&10	&100 &25	  & 0.466 &36.743  & 3.931   & 0.107   \\ \hline
 10	&10	&125 &25	  & 0.530 &39.105  &4.405  & 0.113  \\ \hline
\end{tabular*}
\end{center}
\caption{Comparison of homochiral to heterochiral polymer concentrations in steady state. The top three rows correspond to the parameters of the left plot in Figure \ref{ee_evol}, while the bottom three rows corresponds to the parameters of the right plot. We introduced the ratio of homochiral to heterochiral polymers, ${\cal R} = ({\cal L} + {\cal D})/{\cal M}$. For each of the two groups (top three and bottom rows) note how the ratio is quite insensitive to variations in the reaction rates.}
\label{homohetero_results}
\end{table}

We have performed a detailed statistical study of the reaction equations with fluctuating values for the $L$-- and $D$-- reaction rates $\alpha_L$, $\alpha_D$, $\beta_L$ and $\beta_D$ as described in eqs. \ref{mean} and \ref{rms} scaled by $d$. This involves solving the coupled system of ordinary differential equations above (eqns. \ref{scaledreactioneqs1}--\ref{scaledreactioneqs7}) $N$ times, each with values for the four reaction rates given by $\alpha_L= \bar \alpha+\delta_L$, $\alpha_D= \bar \alpha +\delta_D$, $\beta_L= \bar \beta+\xi_L $, and $\beta = \bar \beta_D+\xi_D$, where the bars denote the mean values (cf. eq. \ref{mean}) and $\delta_{L(D)}$ and $\xi_{L(D)}$ are Gaussian-distributed random numbers (cf. eq. \ref{rms}), within a fixed rms width set by $\alpha_0$ and $\beta_0$.

\begin{figure}[htbp]
\centerline{\includegraphics[width=0.6\linewidth]{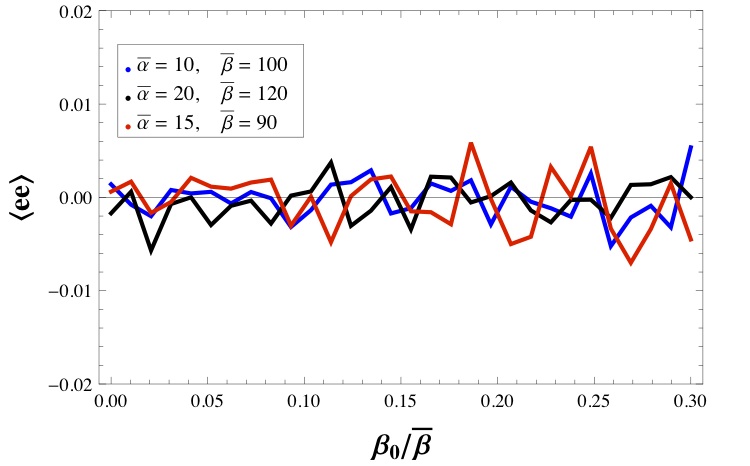}}
\caption{Ensemble averaged enantiomeric excess, $\langle ee \rangle$, for sample sizes $N = 5000$ and fluctuations in $\alpha_{L(D)}$ fixed at $20\%$ ($\alpha_0 = 0.2{\bar\alpha}$), with fluctuations in $\beta_{L(D)}$ varying between $10\%$ and $30\%$ ($0.1 {\bar\beta}\leq \beta_0 \leq 0.3{\bar\beta}$) and fixed $\sigma = 200$. Shown are $\bar{\alpha} = 10$, $\bar{\beta} = 100$ in blue, $\bar{\alpha} = 15$, $\bar{\beta} = 90$ in red, and $\bar{\alpha} = 20$, $\bar{\beta} = 120$ in black.} \label{fig2}
\end{figure}

Our results indicate that the net chirality is overall more sensitive to the amplitude of the chiral-selective variations about the mean values of the reaction rates than to the values of the reaction rates themselves. In Figures 2 and 3 we show the results for three sets of data with variations in $\beta_{L(D)}$ between $10\%$ and $30\%$ ({\it i.e.} $0.1 {\bar\beta}\leq \beta_0 \leq 0.30{\bar\beta}$) and in $\alpha_{L(D)}$ fixed at $20\%$ ({\it i.e.}  $\alpha_0 = 0.2{\bar\alpha}$). Each data point represents an ensemble average over a random sample of $N = 5000$ individual experiments (or prebiotic realizations). Shown are results for $\bar{\alpha} = 10$, $\bar{\beta} = 100$ in blue, $\bar{\alpha} = 15$, $\bar{\beta} = 90$ in red, and $\bar{\alpha} = 20$, $\bar{\beta} = 120$ in black.  Figure \ref{fig2} demonstrates that the ensemble-averaged enantiomeric excess is racemic, with an equal number of systems favoring excesses in $D$ as in $L$, as should be expected. Closer inspection of the details of the distribution demonstrate that some chemistries within the statistical sampling have moderate to large enantiomeric excesses. In Figure 3a, the mean value of the net chirality is shown ({\it i.e.} the mean {\it magnitude} of the enantiomeric excess). Although the distribution of enantiomeric excess is centered on zero as shown in Figure \ref{fig2}, Figure 3a demonstrates that the average net chiral asymmetry obtains moderate values, with the $\langle |ee| \rangle$ varying from $\sim 10\%$ for $\beta_0 = 0.1{\bar\beta}$ to $\sim 16\%$ for $\beta_0 = 0.3{\bar\beta}$, with the $\langle |ee| \rangle$ increasing nearly linearly with the size of the fluctuations. We stress that individual members of the ensemble may have much larger net chiralities than these mean values. The variances of the data set are shown in Figure 3b, and range between $\sim 0.015$ and $\sim0.045$ for $\beta_0 = 0.1{\bar\beta}$ and $\beta_0 = 0.3{\bar\beta}$, respectively. As with the mean value of the net chirality, the variance increases with the amplitude of departure from the mean, leading to a wider spread in the values of the enantiomeric excess with increasing $\beta_0$. For large departures from the mean, say $\beta_0 = 0.3{\bar\beta}$, the variance is $\sim0.04$, corresponding to a standard deviation of $0.2$, indicating that $68\%$ of systems have an $|ee| < 0.2$, and $27\%$ have an enantiomeric excesses in the range $0.2 \leq |ee| \leq 0.4$ (assuming a normal distribution for the data, see Figure 4). A small fraction of $0.2 \%$ of systems possess a sizable $|ee|$ in excess of 0.6. 

In Figure \ref{probability_dist_025}, we further explore the spread in the distribution of the enantiomeric excess in experimental systems for an ensemble with $\bar{\alpha} = 20$, $\bar{\beta} = 120$ and $\alpha_0 = 0.25{\bar\alpha}$ and $\beta_0 = 0.25{\bar\beta}$. Outliers in the distribution have very high enantiomeric excesses up to as much as $80 - 90 \%$, although most systems fall within a range of enantiomeric excesses with $|ee| < 0.25$, where the mean of the distribution lies. In Figure \ref{probability_dist_010} we repeat the analysis with $\bar{\alpha} = 20$, $\bar{\beta} = 100$ and $\alpha_0 = 0.10{\bar \alpha}$ and $\beta_0 = 0.10{\bar\beta}$. Note that even with rms of 10\%, it is still possible to obtain substantial enantiomeric excesses above 20\% at a few percent. Some outliers obtain over 40\% ee.

In Figure \ref{cross_section}, we show a surface plot of the mean values of the net chiral asymmetry for $\beta_{L(D)}$ varying between $10\%$ and $30\%$ ({\it i.e.} $0.1 {\bar\beta}\leq \beta_0 \leq 0.3{\bar\beta}$) and fluctuations in $\alpha_{L(D)}$ varying over the same range. The highest values for the net enantiomeric excess are observed for large fluctuations in $\alpha_{L(D)}$ and $\beta_{L(D)}$, where mean values for the net enantiomeric excess reach as high as $|ee| \sim 0.2$ for the systems under study. These results demonstrate a general trend of increased average $\langle |ee| \rangle$ with increased chiral selectivity in the rates, as well as an increase in the spread of the distribution, indicating that larger departures form the mean lead to a larger fraction of outliers with significant enantiomeric excess.

\begin{figure}[htbp]
\includegraphics[width=0.49\linewidth]{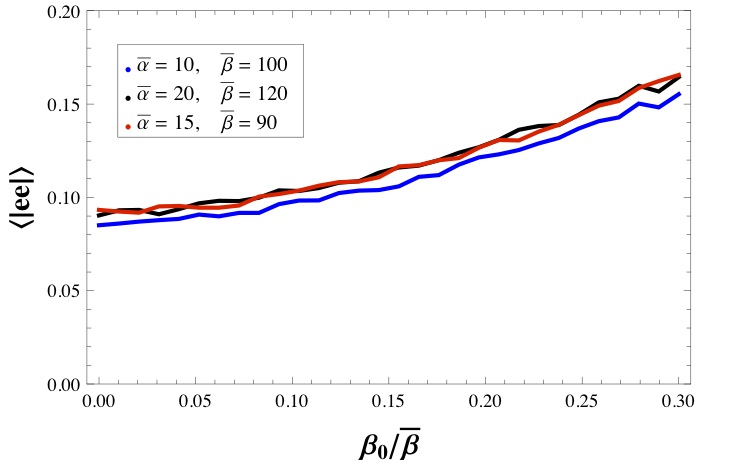}
\includegraphics[width=0.49\linewidth]{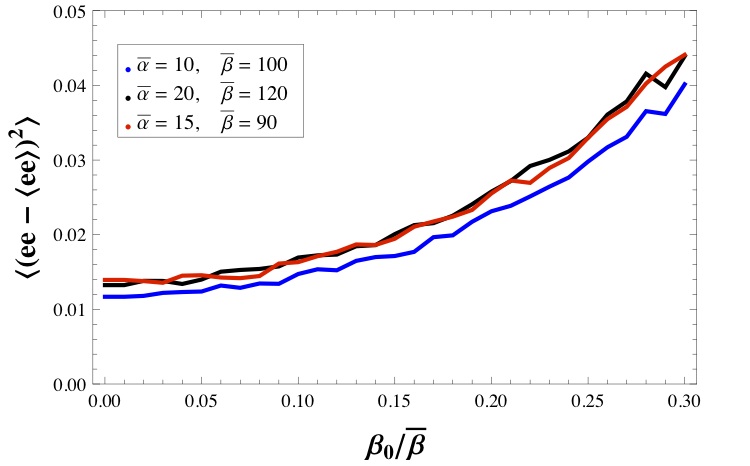} 
\caption{Ensemble averaged mean of the net enantiomeric excess, $|ee|$, and variance for the ensemble data shown in Figure \ref{fig2} with $N = 5000$, with fluctuations in $\alpha_{L(D)}$ fixed at $20\%$ ($\alpha_0 = 0.2{\bar\alpha}$), and fluctuations in $\beta_{L(D)}$ varying between $10\%$ and $30\%$ ($0.1{\bar\beta} \leq \beta_0 \leq 0.3{\bar\beta}$) and fixed $\sigma = 200$. Shown are $\bar{\alpha} = 10$, $\bar{\beta} = 100$ in blue, $\bar{\alpha} = 15$, $\bar{\beta} = 90$ in red, and $\bar{\alpha} = 20$, $\bar{\beta} = 120$ in black. {\bf Left:} Mean values of the net enantiomeric excess, $\langle |ee| \rangle$. {\bf Right:} Variance in the enantiomeric excess $\langle (ee - \langle ee \rangle)^2 \rangle$.}
\label{cross_section}
\end{figure}

\begin{figure}[htbp]
\centerline{\includegraphics[scale=0.75]{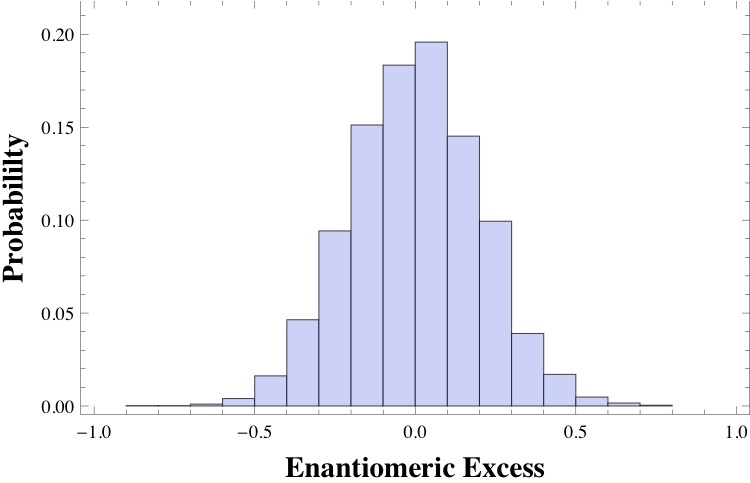}}
\caption{Distribution of ensemble $ee$ values for $N = 5000$ with $\bar{\alpha} = 20$, $\bar{\beta} = 120$, $\alpha_0 = 0.25{\bar\alpha}$ and $\beta_0=0.25{\bar\beta}$, and $\sigma = 200$ .}
\label{probability_dist_025}
\end{figure}

\begin{figure}[htbp]
\centerline{\includegraphics[scale=0.75]{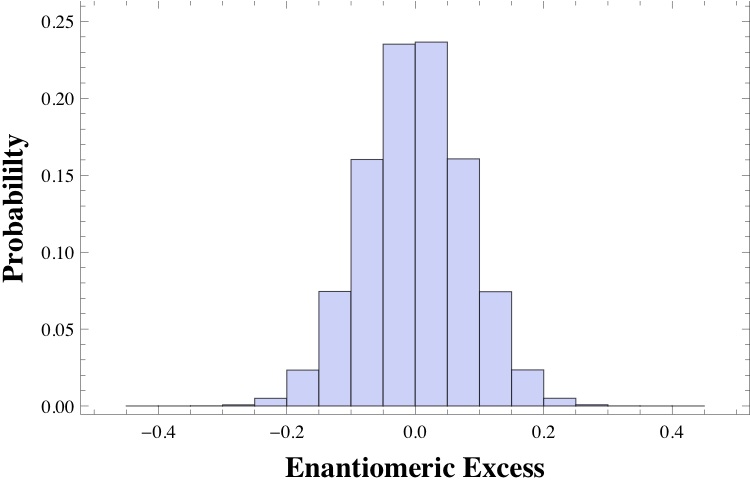}}
\caption{Distribution of ensemble $ee$ values for $N = 1\times 10^6$ with $\bar{\alpha} = 20$, $\bar{\beta} = 100$, $\alpha_0 = 0.10{\bar\alpha}$, $\beta_0 = 0.10{\bar\beta}$, and $\sigma = 200$ .}
\label{probability_dist_010}
\end{figure}

\begin{figure}[htbp]
\centerline{\includegraphics[scale=0.75]{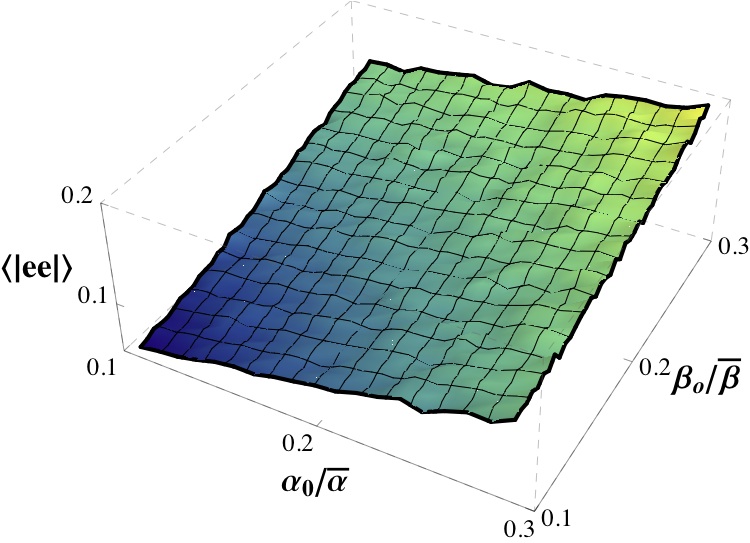}}
\caption{Ensemble averaged net enantiomeric excess, $\langle |ee| \rangle$, for $N = 100$ with fluctuations in $\alpha_{L(D)}$ and $\beta_{L(D)}$ between 10\%  and 30\% ($\alpha_0/{\bar\alpha}$, $\beta_0/{\bar\beta}$ between $0.1$ and $0.3$) with $\bar{\alpha} = 20$, $\bar{\beta} = 200$ and $\sigma = 200$.}
\label{cross_section}
\end{figure}

\section{Summary and Outlook}
\label{sum}

Due to the symmetry between the chemistries of $L$- and $D$- monomers, including the biologically relevant examples of amino acids and nucleotides, it is expected that nearly all chemical interactions between the two enantiomeric forms and their environment are symmetric with respect to both chiralities. However, specific local environmental influences, such as mineral surfaces, chiral-selective temperature sensitivity, photolysis, or the presence of asymmetric catalysts, can bias reactions to favor one chirality over the other to a larger or lesser extent. All of these examples are inherently globally symmetric, such that on average a statistical sampling of chemistries will be racemic ({\it i.e.} some will favor $L$ and others $D$). Asymmetry therefore arises depending on the particular context. For example, surfaces can lead to chiral-selectivity due to differential activity of the two enantiomers when adhered to minerals or clays \citep{Hazen, Ferris}. This can lead specific local chemistries to have a chiral bias in reaction rates that differ significantly from the standard reactivities in free solution. We have explored general scenarios of a similar nature here, within the context of the as yet unknown prebiotic chemistry which may have given rise to the first life-forms. As we have argued, different environmental contexts will lead to varying deviations from the mean reactivity of $L$- and $D$- enantiomers, leading to localized symmetry breaking. Larger fluctuations will result in some chemistries having substantial chiral asymmetries. Based on the results presented here, we see that fluctuations as small as $10\%$ in reactivities of $L$- and $D$- monomers may lead to significant chiral symmetry breaking ({\it i.e.} $|ee| > 10\%$).

Additionally, fluctuations in chiral-selective reaction rates may provide an alternative mechanism to the initiation of chiral asymmetry prior to amplification via autocatalysis with enantiomeric cross-inhibition in a Frank-type mechanism. Traditional approaches assume that fluctuations in the relative numbers of $L$- and $D$- monomers are enough to break the initial symmetry between the two enantiomers, which may then be amplified. However,  fluctuations in chemical systems typically scale as $1/\sqrt(N)$ \citep{Landau}, where $N$ is the number of molecules in the system, and will therefore be exceedingly small for most chemical systems. Deviations from perfect racemates will therefore be correspondingly small and require significant enzymatic enhancement by efficient catalysts to be amplified ({\it e.g.} see \citet{gleiser_walker}). Here we present an alternative and more robust mechanism for symmetry breaking, whereby the initial symmetry breaking occurs at a macroscopic rather than microscopic level, being a manifestation of fluctuations in reactivities, rather than in molecular numbers. Such scenarios obviate the need for special asymmetric initial conditions, allowing asymmetry to emerge as the reactions unfold, due to asymmetries present in the local chemistry. 

Finally, we note that our results suggest that the origin of homochirality on the prebiotic Earth was a statistical process, potentially hosting many abiogenic events, few of which were highly asymmetric. Quite likely, only those chemistries which were highly asymmetric were capable of giving rise to life \citep{chiralreview1,chiralreview2}, and only one of these trials ultimately led to the emergence of the LUCA. This has different observational consequences for experimental systems as well as searches for biosignatures on other worlds than the exogenous or endogenous sources of chiral excess described above. Parity-violation would result in a universal bias toward a particular chiral asymmetry, while UV-irradiation might lead to chiral biases on the scale of planetary systems. Here, chiral selectivity is a local statistical phenomenon, and we should expect a large sampling of extraterrestrial stereochemistries to be racemic on average.

\begin{acknowledgements}

MG is supported in part by a National Science Foundation grant PHY-1068027. BJN is a Presidential Scholar at Dartmouth College. SIW gratefully acknowledges support from the NASA Astrobiology Institute through the NASA Postdoctoral Fellowship Program.

\end{acknowledgements}

\end{document}